\documentclass[useAMS,usenatbib,usegraphicx]{mn2e}
\include{epsf}
\newcommand\aj{{AJ}}%
\newcommand\araa{{ARA\&A}}%
\newcommand\apj{{ApJ}}%
\newcommand\apjl{{ApJ}}%
\newcommand\apjs{{ApJS}}%
\newcommand\aap{{A\&A}}%
\newcommand\mnras{{MNRAS}}%
\newcommand\na{{New A}}%
\newcommand\pasp{{PASP}}%
\newcommand\pasj{{PASJ}}%
\title[Gas and Stellar Motion of Co-Rotating Spiral Arms]
{Gas and Stellar Motions and Observational Signatures of Co-Rotating Spiral Arms}
\author[Kawata et~al.]
 {\parbox{\textwidth}{Daisuke~Kawata$^{1}$\thanks{E-mail: d.kawata@ucl.ac.uk},
Jason~A.~S.~Hunt$^{1}$, Robert~J.~J.~Grand$^{1}$, Stefano Pasetto$^{1}$, Mark Cropper$^{1}$}\vspace{0.5cm}
\\
$^{1}$ Mullard Space Science Laboratory, University College London,
Holmbury St. Mary, Dorking, Surrey, RH5 6NT, UK
}
\date{Accepted .
      Received ;
      in original form }

\pagerange{\pageref{firstpage}--\pageref{lastpage}}
\pubyear{2013}

\begin{document}

\maketitle

\label{firstpage}

\begin{abstract}
We have observed a snapshot of our N-body/Smoothed Particle Hydrodynamics simulation of a Milky Way-sized barred spiral galaxy in a similar way to how we can observe the Milky Way. The simulated galaxy shows a co-rotating spiral arm, i.e. the spiral arm rotates with the same speed as the circular speed. We observed the rotation and radial velocities of the gas and stars as a function of the distance from our assumed location of the observer at the three lines of sight on the disc plane, $(l,b)=(90,0), (120,0)$ and (150,0) deg. We find that the stars tend to rotate slower (faster) behind (at the front of) the spiral arm and move outward (inward), because of the radial migration. However, because of their epicycle motion, we see a variation of rotation and radial velocities around the spiral arm. On the other hand, the cold gas component shows a clearer trend of rotating slower (faster) and moving outward (inward) behind (at the front of) the spiral arm, because of the radial migration. We have compared the results with the velocity of the maser sources from \citet{rmbzd14}, and find that the observational data show a similar trend in the rotation velocity around the expected position of the spiral arm at $l=120$ deg. We also compared the distribution of the radial velocity from the local standard of the rest, $V_{\rm LSR}$, with the APOGEE data at $l=90$ deg as an example.
\end{abstract}


\section{Introduction}
\label{intro-sec}

 Spiral arm structures are the beautiful structures that have fascinated astronomers for a long time. One of the long-time mysteries of the spiral arm was the so-called ``winding-dilemma''. From the observations of the rotation curve of disc galaxies, it is known that the stars in the inner region of the disc rotate faster, i.e. the angular speed is higher, than the stars in the outer region. Therefore, if the spiral arm is a material arm, i.e. the spiral arm is moving at the same speed as the stars, the spiral arm should wind up quickly \citep[e.g.][]{ejw896}. Spiral density wave theory described in \citet{ls64} solved the issue by considering the spiral pattern as a density wave. Then, the density wave can be a rigidly rotating feature with a constant pattern speed, irrespective of the stellar rotation speed, and consequently long-lived.
 
Recently, thanks to the powerful computational facilities, the resolution of three-dimensional N-body simulations has improved significantly, and the artificial heating from which the previous low-resolution simulations suffered is minimized \citep[e.g.][]{fbsmk11,jas13}. Such high-resolution simulations allow us to study further the spiral arm theory. However, even with such high-resolution simulations, so far no single N-body simulation reproduces a long-standing spiral arm feature such as that is suggested in \citet{ls64} \citep{jas11}. Recent studies show that the spiral arms in the numerical simulations are transient and recurrent \citep[e.g.][]{cldb06,wbs11,gkc12a,gkc12b,rdqw12,bsw13,dvh13,rvfrv13}. For example, \citet{gkc12a,gkc12b} demonstrated that the spiral arm was rotating with the same speed as the stars, i.e. co-rotating \citep[see also][]{wbs11}, and therefore winding. Still in each snapshot, the spiral arms are always apparent, and the spiral arms are constantly forming and disrupting, i.e. recurrent, with a lifetime of about 100 Myr. Although the co-rotating spiral arm leads to the winding-dilemma, \citet{gkc13} demonstrated that the spiral arms were disrupted before they wound up completely, and the pitch angle of the spiral arms correlated with the shear rate of the disc, as observed \citep[e.g.][]{sbbh06}. Interestingly, the winding nature of the spiral arm seen in N-body simulations can naturally explain the observed scatter in the correlation between the pitch angle and the shear rate \citep[see][for more thorough discussion]{gkc13}.

\citet{gkc12a} demonstrated that the spiral arms in N-body simulations were forming with a similar mechanism to the so-called swing amplification theory suggested by \citet{jt66} and \citet{at81} \citep[see also][]{bsw13,dvh13}. However, while the swing amplification is considered to happen at a single co-rotation radius, where the spiral arm pattern speed is consistent with the rotation speed of the stars, in the numerical simulations the co-rotation resonance occurs at all radii, and the swing amplification is happening (or propagating) at every radius. This is one of the current explanations for the transient and winding spiral arm features. In others, for example, the transient features of the spiral arm can be interpreted as overlapping multiple-wave modes \citep[e.g.][]{qdbmc11,rdqw12,sc14}, where each mode appears around the co-rotation radius and is relatively long lived. Still, there is no clear explanation of the origin and nature of the spiral arms, which remains as a challenge for the galactic astronomer. 

Interestingly, there is also observational evidence against long-lived spiral arms. For example, \citet{mrm06} analysed the pattern speed of the spiral arm as a function of radius for NGC 1068 using their generalized version of the method from \citet{tw84}, so-called Tremane-Weinberg method. They showed that the pattern speed of the spiral arm decreases with radius, and therefore the lifetime of the spiral arm must be short. Similar pattern speeds were also observed with the same technique in other galaxies \citep[e.g.][]{mrmds08,mrmsv08,mrm09,sw11,sw12}. However, the accuracy and validity of the Tremaine-Weinberg method are still needed to be tested against the future observations \citep[e.g.][]{mrmds08,rvfrv13}. Another observational test is the so-called ``offset'' method. If there is a long-lived rigidly rotating spiral arm, one can define the co-rotation radius. Since the angular velocity of gas and stars is observed to be faster in the inner region, the gas and stars in the region inside (outside) the co-rotation radius will move faster (slower) than the spiral arm. The gas component piles up in the spiral arms, experiencing a shock that induces star formation \citep{mf68,wr69}. In this scenario, the youngest stars born from the molecular clouds in the spiral arms would be found slightly ahead of the arm traced by the molecular gas, if located within the co-rotation radius, and behind the arm outside of the co-rotation radius. Therefore, if we observe the tracers of the different stages of star formation, such as HI, CO and H$\alpha$, an offset among them as a function of radius would be expected. By combining H$\alpha$ imaging and $Swift/UVOT$ Near-Ultraviolet (NUV) data, \citet{fckph12} distinguished the regions with ongoing star formation and the regions with star formation a few hundred million years ago in the grand-design spiral galaxy, M100. Contrary to the expectation from the density-wave theory, no offset was found between these two regions. The same conclusion was reached in \citet{frdlw11}, although some studies claimed to find a significant offset for some galaxies \citep[e.g.][]{eksnk09,trwbd08,mggl13,hkbeh14}. 

The Milky Way is a (barred) spiral galaxy \citep[e.g][]{ds01,bcbim05,jpv13} which we can observe in great detail. For example, the detailed map of HI and CO observations provide the global position and kinematics of the spiral arm in the gaseous phase \citep[e.g.][]{dht01,ns03,ns06,kk09}. Star clusters provide us with reasonable photometric distances, and young star clusters can be used to trace the spiral arm and also measure the pattern speed of the spiral arm using a similar technique to the offset method mentioned above \citep[e.g.][]{dl05,ns07}. The influence of the spiral arms on the stellar motion has also been measured and compared with the models \citep[e.g.][]{fft01,avpmff09,jas10,sfbbf12,fsf14}. The maser sources associated with high-mass star forming regions are recognized as a unique source to trace the spiral arm structures, because Very Long Baseline Interferometry (VLBI) observations allow their parallaxes and proper motions to be measured with great accuracy, $\sim10$ $\mu$as. For example, recently the Bar and Spiral Structure Legacy (BeSSeL) Survey and Japanese VLBI Exploration of Radio Astrometry (VERA) provided the parallaxes and proper motion measurements for over 100 maser sources \citep{rmbzd14}. In the future, the European Space Agency's {\it Gaia} satellite (launched in December 2013) will produce accurate measurements of the parallax and proper motion for about a half billion disc stars \citep[e.g.][]{rlrig12}. 

To study the nature of the spiral arms from these observations, we need to compare the observational data with the theoretical prediction from different scenarios of the origins of spiral arms. Although there are many successful studies for reconstructing the Galactic bar structure and the pattern speed by comparing the observational data with the theoretical model prediction, it is more complicated for the spiral arms \citep[e.g. see][for a review]{og11}. \citet{mq08} assumed the rigid rotation of the spiral arm, and made predictions of the radial velocity distribution of stars in the cases of different number of arms, pitch angles and pattern speed.  \citet{afrpvm11} studied how the rigidly-rotating spiral structures affect the stellar kinematics and the distribution of radial and rotational velocities of the stars \citep[see also][]{qdbmc11}. \citet{rafvrp14} studied the vertex deviation map from both rigidly-rotating spiral arms and transient spiral arms. \citet{bammsw09} discussed the origin of the large peculiar velocities observed for the maser sources at that time \citep{rmzbm09}. They argued that such large peculiar velocities are difficult to be explained with the density wave theory, but using N-body/Smoothed Particle Hydrodynamics (SPH) simulations, they showed that the observed large peculiar velocities of the maser sources can be reproduced by the transient and recurrent spiral arms. 
Therefore, more predictions for the observable signatures from the transient and recurrent spiral arms in the numerical simulations should be valuable for future and on-going observational surveys. 

In this paper, we study the kinematic structure of both the star and gas components around a spiral arm in a simulated barred galaxy similar in size to the Milky Way. The simulated galaxy has a transient, recurrent and co-rotating spiral arm similar to that seen in \citet{gkc12b}. We target a spiral arm similar to the Perseus arm. We then make a prediction of the observational signatures of the kinematics of the stars and gas around the Perseus arm, if it is also a transient, recurrent and co-rotating spiral arm. 

Section~\ref{meth-sec} describes briefly the numerical simulation code and numerical models.
Section~\ref{res-sec} presents the results.
A summary of this study is presented in Section~\ref{sum-sec}.

\section{Method and Models}
\label{meth-sec}

\subsection{Numerical Simulation code}
\label{code-sec}

We use our original N-body/SPH \citep{ll77,gm77} code, {\tt GCD+}, which can be used in studies of galaxy formation and evolution in both a cosmological and isolated setting \citep{kg03a,bkw12,kogbc13,kgbgr14}.
{\tt GCD+} incorporates self-gravity, hydrodynamics, radiative cooling, star formation, SNe feedback, and metal enrichment, including metal diffusion \citep{ggb09}.
We have implemented a modern scheme of SPH suggested by \citet{rp07}, including their artificial viscosity switch \citep{jm97} and artificial thermal conductivity to resolve the Kelvin-Helmholtz instability \citep[see also][]{rh12,pfh13,sm13,hnwmo14}. Following \citet{sh02}, we integrate the entropy equation instead of the energy equation. As suggested by \citet{sm09}, we have added the individual time step limiter, which is crucial for correctly resolving the expansion bubbles induced by SNe feedback \citep[see also][]{mbg10,dd12}. We also implement the FAST scheme \citep{sm10} which allows the use of different timesteps for integrating hydrodynamics and gravity. The code also includes adaptive softening length for the gas particles \citep{pm07}. Our recent updates and performance in various test problems are presented in \citet{bkw12}  and \citet{kogbc13}.

 Radiative cooling and heating are calculated with CLOUDY \citep[v08.00:][]{fkv98} following \citet{rk08}. We tabulate cooling and heating rates and the mean molecular weight as a function of redshift, metallicity, density and temperature adopting the 2005 version of the \citet{hm96} UV background radiation. The details of the recipe of star formation and stellar feedback are described in \citet{kgbgr14}. We adopt the star formation threshold density, $n_{\rm H,th}=20$ cm$^{-3}$.  The main free parameters of our feedback model include energy per SN, $E_{\rm SN}$, and stellar wind energy per massive star, $L_{\rm SW}$, and we adopt $E_{\rm SN}=10^{50}$ erg and $L_{\rm SW}=10^{36}$ erg~s$^{-1}$ for this paper. The other parameters are the same as \citet{kgbgr14}.

\subsection{Milky Way Sized Disc Simulation}
\label{inic-sec}

 We simulate the evolution of a barred disc galaxy similar in size to the Milky Way. We initially set up an isolated disc galaxy which consists of gas and stellar discs, with no bulge component, in a static dark matter halo potential \citep{rk12,gkc12b}, to save computational costs. A live dark matter halo can respond to the disc particles by exchanging angular momentum. This becomes important for long term evolution of a bar \citep[e.g.][]{ds00,ea02,ea03}. However, the effect of the assumed static dark matter halo is expected to be small for transient spiral arms which we focus on in this paper. Furthermore, a live dark matter halo is often modeled with particles more massive than disc particles, which may introduce some scattering and heating \citep[e.g.][]{dvh13}. Therefore, in the interest of computational speed and a more controlled experiment, we use a static dark matter halo.
 We use the standard Navarro-Frenk-White (NFW) dark matter halo density profile \citep{nfw97}, assuming  a $\Lambda$-dominated cold dark matter ($\Lambda$CDM) cosmological model with cosmological parameters of $\Omega_0=0.266=1-\Omega_{\Lambda}$, $\Omega_{\rm b}=0.044$, and $H_0=71{\rm kms^{-1}Mpc^{-1}}$:

\begin{equation}
\rho _{dm}=\frac{3H_{0}^{2}}{8\pi G}\frac{\Omega _{0}-\Omega_b}{\Omega_0}\frac{\delta _{c}}{cx(1+cx)^{2}},
\end{equation}
where
\begin{equation}
c=\frac{r_{200}}{r_{s}}, \;\; x=\frac{r}{r_{200}},
\end{equation}
and
\begin{equation}
r_{200}=1.63\times 10^{-2}\left(\frac{M_{200}}{h^{-1}M_{\odot }}\right)^{\frac{1}{3}} h^{-1}\textup{kpc},
\end{equation}
\noindent where $\delta _{c}$ is the characteristic density of the profile \citep{nfw97}, $r$ is the distance from the centre of the halo and $r_{s}$ is the scale radius. The total halo mass is set to be $M_{200}=2.5\times 10^{12}M_{\odot }$ and the concentration parameter is set at $c=10$. The halo mass is roughly consistent with (or slightly higher than) the recent measured mass of the Milky Way \citep[e.g.][]{pjm11}. 

The stellar disc is assumed to follow an exponential surface density profile:
\begin{equation}
\rho _{d,*}=\frac{M_{d,*}}{4\pi z_{d,*}R_{d,*}^2}\textup{sech}^{2}\left(\frac{z}{z_{d,*}}\right)\exp \left(-{\frac{R}{R_{d,*}}}\right),
\end{equation}
\noindent where $M _{d,*}$ is the stellar disc mass, $R_{d,*}$ is the scale length and $z_{d,*}$ is the scale height. Following the observational estimates of the Milky Way, we adopt $M _{d,*}=4.0\times10^{10}$ M$_{\odot}$, $R_{d,*}=2.5$ kpc and $z_{d,*}=0.35$ kpc. For simplicity, we set up only a thin disc component, and ignore the thick disc contribution. 

The gaseous disc is set up following the method described in \citet{sdh05b}.
The radial surface density profile is assumed to follow an exponential law like the stellar disc with a scale length, $R_{d,gas}=8.0$ kpc. The initial vertical distribution of the gas is iteratively calculated to reach hydrostatic equilibrium assuming the equation of state calculated from our assumed cooling and heating function. The total gas mass is $1.0\times 10^{10}$ M$_{\odot }$. 

 We use 1,000,000 gas particles and 4,000,000 star particles for the initial condition. This leads to 10,000 M$_{\odot}$ for each particle, and means that both the star and gas particles have the same mass resolution in our simulations. This is required from our modelling of star formation, feedback and metal diffusion. We apply the spline softening and variable softening length suggested by \citet{pm07} for SPH particles. We set the minimum softening length at 158 pc (the equivalent Plummer softening length is about 53 pc), which corresponds to the required softening for gas to resolve $n_{\rm H}=1$ cm$^{-3}$ with Solar metallicity.
 
\section{Results}
\label{res-sec}

\begin{figure}
\centering
\includegraphics[width=\hsize]{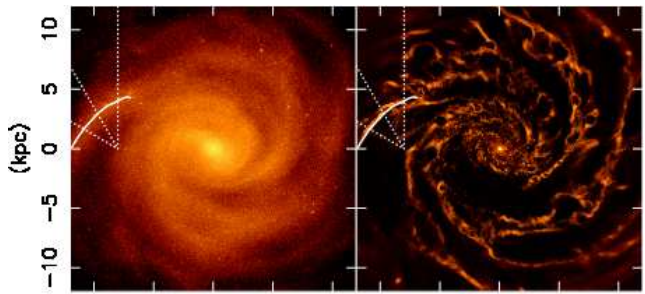}
\caption{
 Snapshot of the simulated galaxy which is used in this paper. Left (Right) panel shows the face-on view of the star (gas) particle distribution. The solid line indicates the position of the spiral arm identified. The observer is assumed to be located at $(x,y)=(-8, 0)$ kpc. Three line of sight directions ($l_{\rm LOS}=90, 120$ and 150 deg) are highlighted with the dotted lines. The galaxy is rotating clockwise. 
}
\label{snap-fig}
\end{figure}

\begin{figure}
\centering
\includegraphics[width=\hsize]{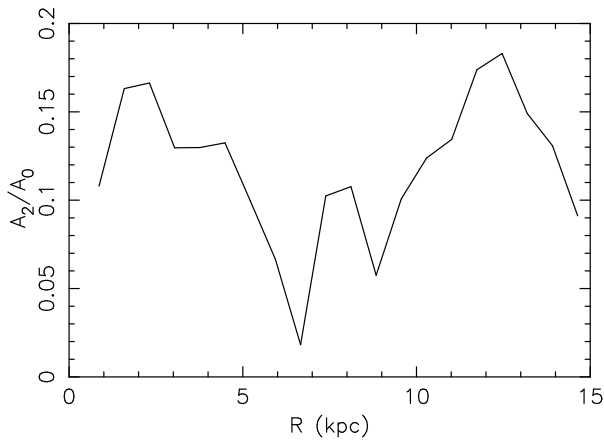}
\caption{
Amplitude of the $m=2$ Fourier mode normalised to the $m=0$ mode as a function of radius.
}
\label{ampm2-fig}
\end{figure}

  We ran a simulation for 1 Gyr from the initial conditions described above. We do not include any continuous external inflow of gas for the simplicity of the numerical setup. We have chosen a snapshot at $t=0.925$ Gyr, because it shows a spiral arm in a similar location to the Perseus arm located in the second quadrant of the Milky Way. The snapshot is shown in Fig.~\ref{snap-fig}. We assumed the location of the observer at $(x,y)=(-8,0)$ kpc in the simulated galaxy. The mean circular velocity at the radius of 8 kpc is $V_{\rm circ,sim}=228$ km~s$^{-1}$. In this paper, we focus on the spiral arm highlighted in Fig.~\ref{snap-fig}. The pitch angle of the spiral arm is 39 deg, which is much larger than the typically estimated pitch angle for the Milky Way spiral arms, $\sim 12$ deg \citep[e.g.][]{jpv14}. Fig.~\ref{ampm2-fig} shows the amplitude of the $m=2$ Fourier mode normalised to the $m=0$ mode calculated with equation (12) of \citet{gkc13}.  The amplitude is similar to the observed spiral galaxies \citep[e.g.][]{rz95}. We focus on the spiral arm in the outer region, $R>8$ kpc, because the amplitude of the spiral arm is higher and the spiral arm is clearer in the outer region. Note that the simulation used is not intended to reproduce the whole structure of the Milky Way. The aim of this study is to qualitatively discuss the kinematical signatures of the co-rotating spiral arm in general.
  
\begin{figure}
\centering
\includegraphics[width=\hsize]{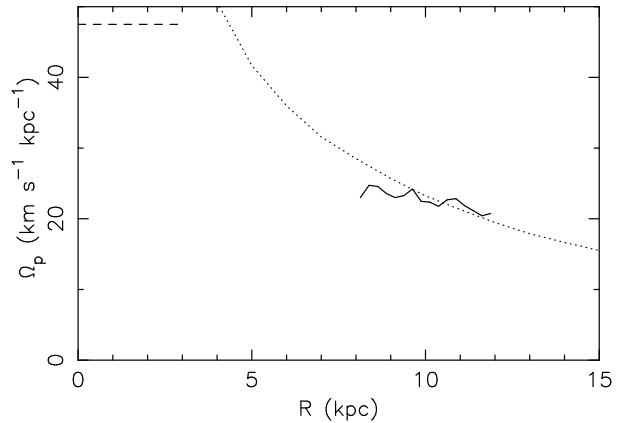}
\caption{
 The measured pattern speed of the spiral arm (solid line), and the circular velocity of the simulated galaxy (dotted line) as a function of the galactocentric radii. The dashed line shows the bar pattern speed for a reference.
}
\label{omgpr-fig}
\end{figure}

 Fig~\ref{omgpr-fig} shows the measured pattern speed of the spiral arm. The pattern speed is measured in the same way as \citet{gkc12b} by tracing the density peak. Though the pattern speed shows some scatter. The pattern speed looks flat, but with a slight trend of decreasing pattern speed with the radius. The pattern speed is similar to the circular speed in the radial range between 8 and 12 kpc. This means that the spiral arm is co-rotating with the stars. We have confirmed that the spiral arm slowly winds up, and it is disrupted at $t\sim0.98$ Gyr, i.e. transient.
 
Fig.~\ref{omgpr-fig} also shows the pattern speed of the bar. The pattern speed of the bar is measured with spectrogram analysis of the $m=2$ mode as described in \citet{gkc12a}, because the density structure of the bar keeps changing and it is difficult to define the bar position angle for each snapshot \citep[see also][]{mvg11,rgaaf11}. Using a gravitational field method \citep[e.g.][]{bb01,bvsl05} described in \citet{gkc12b}, we obtain the bar strength of $Q_b=0.15$. Table 1 of \citet{rgaaf11} summarises the current estimates of the strength of the bar in the Milky Way, which is between $Q_b=0.17$ and 0.83. The bar strength of our simulated galaxy is consistent with the lowest estimate for the bar strength of the Milky Way. \citet{gkc12b} showed numerical simulations with a bar of $Q_b=0.27$ and 0.11. \citet{gkc12b} demonstrated that although the stronger bar causes a flatter pattern speed of the spiral arm, compared with simulations without a bar, the pattern speed of the spiral arm is still similar to the rotation velocity of stars in the spiral galaxy with the bar of this level of strength. \citet{rvfrv13} also suggested that the strong influence of the bar can induce more rigidly rotating spiral arm, which could be tested with the Galactic disc surveys, such as {\it Gaia} and APOGEE.

 We now show the kinematic features expected from the co-rotating spiral arm. We focus on the properties of the star and gas particles within the lines of sight highlighted by the white dotted lines in Fig.~\ref{snap-fig}. These lines of sight correspond to the galactic latitude of $b_{\rm LOS}=0$ and galactic longitudes of $l_{\rm LOS}=90, 120$ and 150 deg. 

\begin{figure*}
\centering
\includegraphics[width=\hsize]{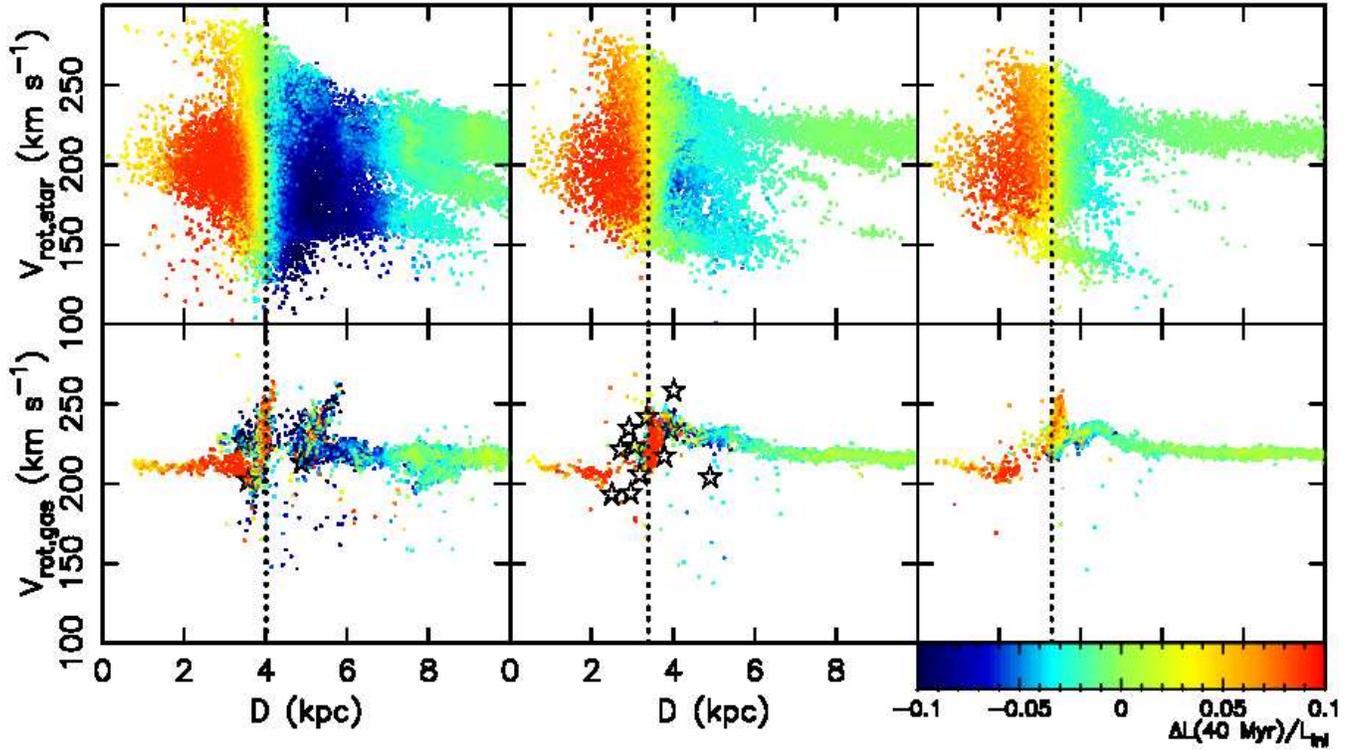}
\caption{
 Rotation velocity of stars (upper panels) and gas (lower panels) as a function of the distance from the observer. Left, middle and right panels show the sample of stars and gas particles within the galactic latitude range of $-5<b<5$ deg and the galactic longitude range of $85<l<95$, $115<l<125$ and $145<l<155$ deg, respectively. The vertical dotted lines indicate the position of the spiral arm highlighted in Fig.~\ref{snap-fig}. Colours correspond to the gain (redder) or loss (bluer) of the angular momentum from 20 Myr ago to 20 Myr later as indicated at the bottom-right corner. The star symbols in the lower panels are the rotation velocity of the maser sources in the Milky Way \citep{rmbzd14}. The rotation velocities are scaled by the difference in the circular velocities of the simulated galaxy and the Milky Way whose observed value we adopt is $V_{\rm circ}=240$ km~s$^{-1}$ \citep{rmbzd14}.
}
\label{disvrot-fig}
\end{figure*}

\begin{figure*}
\centering
\includegraphics[width=\hsize]{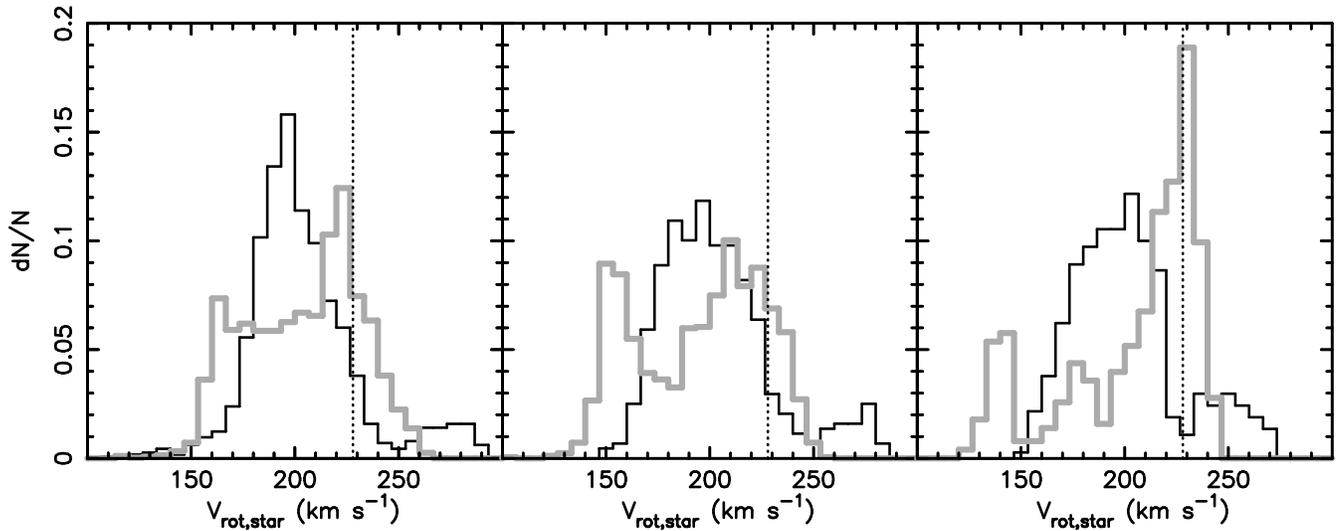}
\caption{
Distribution of rotation velocity of stars whose distance range is $D_{\rm sp}-2<D<D_{\rm sp}-1$ kpc (black, trailing side) and $D_{\rm sp}+1<D<D_{\rm sp}+2$ kpc (grey, leading side), where $D_{\rm sp}$ is the distance of the spiral arm (dotted lines in Fig.~\ref{disvrot-fig}). Left, middle and right panels show the rotation velocity distribution for stars within the galactic latitude range of $-5<b<5$ deg and the galactic longitude range of $85<l<95$, $115<l<125$ and $145<l<155$ deg, respectively. The vertical dotted lines indicate the circular velocity of the simulated galaxy, i.e. $V_{\rm circ,sim}=228$ km s$^{-1}$.
}
\label{vrothis-fig}
\end{figure*}

\begin{figure*}
\centering
\includegraphics[width=\hsize]{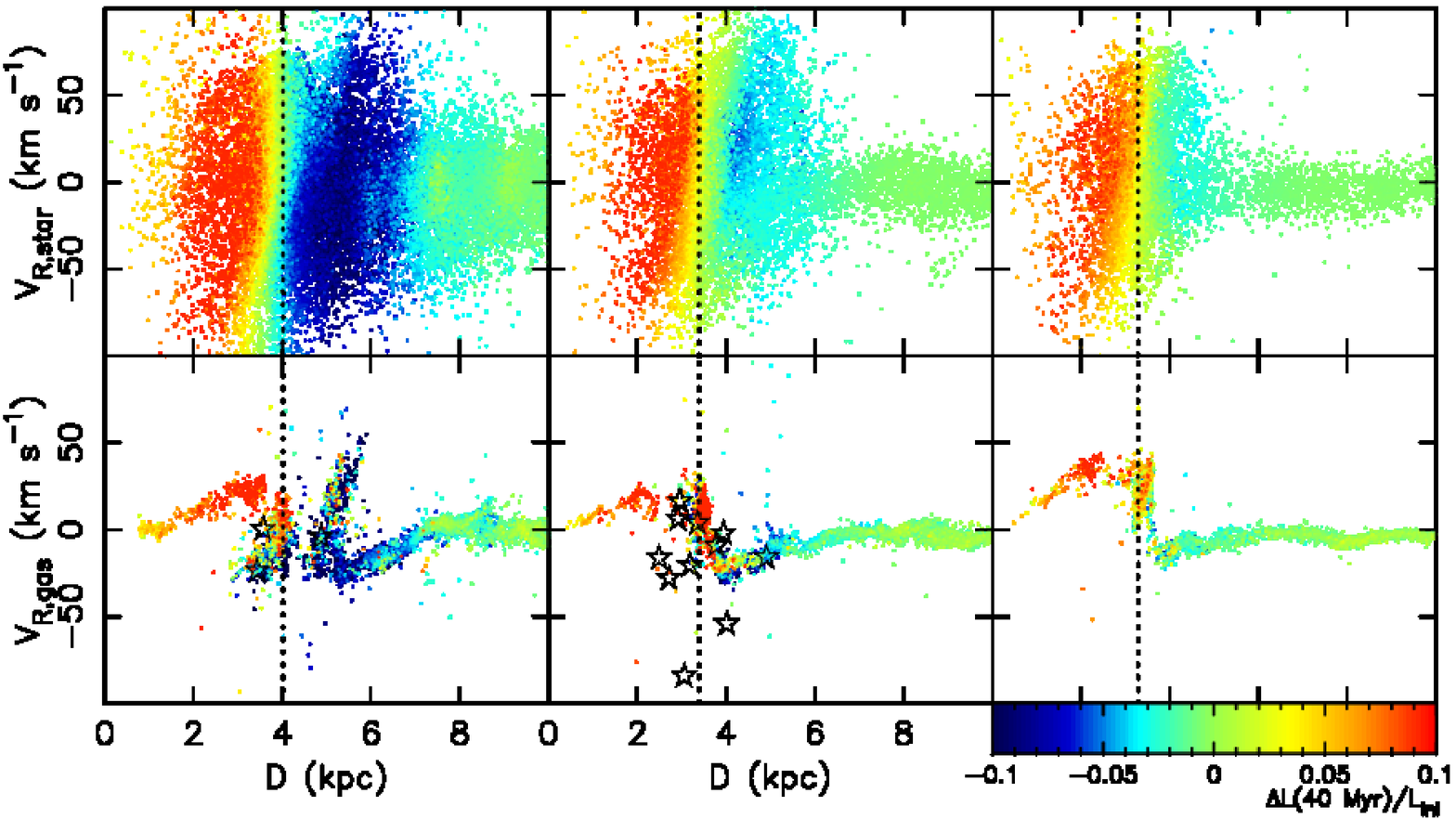}
\caption{
 Galactocentric radial velocity of stars (upper panels) and gas (lower panels) as a function of the distance from the observer. Left, middle and right panels show the sample of stars and gas particles within the galactic latitude range of $-5<b<5$ deg and the galactic longitude range of $85<l<95$, $115<l<125$ and $145<l<155$ deg, respectively. The vertical dotted lines indicate the position of the spiral arm highlighted in Fig.~\ref{snap-fig}.  Colours correspond to the gain (redder) or loss (bluer) of the angular momentum from 20 Myr ago to 20 Myr later as indicated at the bottom-right corner. The star symbols in the lower panels are the Galactocentric radial velocity of the maser sources in the Milky Way \citep{rmbzd14}. 
}
\label{disvr-fig}
\end{figure*}

The upper panels of Fig.~\ref{disvrot-fig} demonstrate the rotation velocity of the stars as a function of their distance from the observer. The sample of star particles are selected within a square angular range of $b_{\rm LOS}\pm5$ and $l_{\rm LOS}\pm5$ deg from the three lines of sight highlighted in Fig.~\ref{snap-fig}. The vertical line in each panel shows the position of the spiral arm, whose galactocentric radius corresponds to $R_{\rm G}=8.9, 10.1$ and 10.9 kpc \footnote{The mean circular velocity slightly increases with the radius in this radial range. The circular velocity is about 234 km~s$^{-1}$ at 11 kpc. Because the difference is small, we use $V_{\rm circ,sim}=228$ km~s$^{-1}$ irrespective of the galactocentric radius.}. Note that the left (right) hand side of the vertical line, i.e. closer to (further from) the observer, is the trailing (leading) side of the spiral arm. The colours of the dots indicate the angular momentum change of the star particles during the $\pm20$ Myr evolution from the time step that we focus on. The short time window of $\Delta t=$40~Myr is chosen to highlight the angular momentum change at the snapshot. The stars in the trailing (leading) side of the spiral arm tend to gain (lose) angular momentum. This trend is much clearer in the edge of the arm, compared to around the centre of the arm where there are stars gaining (losing) angular momentum at the slightly leading (trailing) side of the spiral arm. \citet{gkc14} showed that the angular momentum loss and gain was due to the tangential force from the spiral arm, and the stars at the leading (trailing) side lost (gained) angular momentum at each time. The tangential force is highest at the edge of the spiral arm, and changes the sign at the centre of the spiral arm.   As a results, the radial migration due to the spiral arm is more effective at the edge of the spiral arm than around the centre of the spiral arm. Also note that, the position of the stars in Fig.~\ref{disvrot-fig} is at the snapshot we chose, but the angular momentum gain and loss is measured within the time bin of $\Delta t=$40~Myr. Stars around the centre of the spiral arm can easily move to the other side of the spiral arm within 40 Myr, which blurs the trend of the loss and gain of angular momentum with respect to the side of the spiral arm. Radial migration (change of the angular momentum) is happening at all radii, because the spiral arm is co-rotating with the stars, which is consistent with \citet{gkc12b,gkc12a}.

  A large range of rotation velocities are seen around the spiral arm. Fig.~\ref{vrothis-fig} compares the distribution of the rotation velocity in the trailing side and the leading side, where we selected the star particles in the distance range between 1 and 2 kpc away from the location of the spiral arm (the solid line in Fig.~\ref{snap-fig}) at each line of sight. As shown in \citet{gkc14}, in general, stars in the trailing (leading) side rotate slower (faster), because they tend to be apo-centre (peri-centre) phase. Still, there are some stars which rotate very fast (slow), but are in the trailing (leading) side of the spiral arm. Especially in the leading side, the rotation velocity distribution shows (at least) two peaks. Although one of the peaks is faster than the peak of the rotation distribution in the trailing side and close to the circular velocity, the other peak is slower than the peak in the trailing side. 
  
  The slowly rotating stars in the leading side are also visible in Fig.~\ref{disvrot-fig}. Following \citet{gkc14}, we identified stars with V$_{\rm rot}\sim$165 km~s$^{-1}$ in the leading side (D$\sim$8 kpc) at $l_{\rm LOS}=90$ deg and tracked their evolution. They were located at an inner radius very close to the arm, but slightly in the trailing side, when the spiral arm was forming. They were at the peri-centre phase then, and were accelerated by the spiral arm formation. Consequently, they overshoot the spiral arm, and at the time of the snapshot they are at the apo-centre phase and are decelerated by the arm, which leads to their slow rotation.  

     We also tracked the stars rotating faster than 250 km~s$^{-1}$ in the trailing side (D$\sim$3 kpc) of the arm at $l_{\rm LOS}=90$ deg. They were rotating at a larger radius before the spiral arm formed. They were slightly in the leading side of the arm, when the spiral arm starts forming. Then, they were passed by the spiral arm, i.e. moved from the leading side to the trailing side, just before the time step selected. They are close to the peri-centre phase at the selected time step. These particles will pass the spiral arm, i.e. they will move from the trailing side to the leading side, after this time step. Therefore, the particles are orbiting around the spiral arm. The particles that have this type of the orbit are called ``non-migrators'' in \citet{gkc14}.
   
  The lower panels of Fig.~\ref{disvrot-fig} show the rotation velocity of the gas component as a function of distance from the observer in the same line-of-slight as the upper panels. Because the gas component does not have the epicyclic motion which leads to the variation in the rotation velocities for stars, there is a clear trend seen around the spiral arm that the gas rotates slower in the trailing side and faster in the leading side, especially at $l_{\rm LOS}=120$ and 150 deg. (As seen in Fig.~\ref{snap-fig} at $l_{\rm LOS}=90$ deg, there is a bubble induced by SNe, which disturbs the kinematics of the gas component.) We consider this to be the result of the radial migration of the gas. For example, the gas in the trailing side gains angular momentum and moves to larger radii. Because the spiral arm is trailing, if the migrating gas is accelerated up to the same speed as the gas at the new radius, it will pass the centre of the arm. To stay in the trailing side at the new outer radius, it should be accelerated to a slightly slower rotation velocity than that of the spiral arm. The opposite is required in the leading side, and only the gas rotating faster can stay in the leading side of the arm.   
   
 These velocity features in stars and gas are due to the co-rotation of the spiral arm. These features are seen at different galactic longitude samples, because the spiral arm is co-rotating at all of the radial range on which we focus. These features, therefore, would be good observable indicators to examine if the Milky Way spiral arms are co-rotating arms.
 
 Recently, \citet{rmbzd14} published their parallax and proper motion measurements for over 100 of the maser sources. We converted their proper motion to the rotation velocity, assuming that the Sun's rotation velocity with respect to the circular velocity is $V_{\odot}=14.6$ km s$^{-1}$ and the Galactocentric distance of the Sun is $R_{\rm G,\odot}=8.34$~kpc \citep{rmbzd14}. We also scaled the rotation velocity to adjust from the measured circular velocity at the Solar radius of $V_{\rm circ}=240$~km~s$^{-1}$ \citep{rmbzd14} to the circular velocity of the simulated galaxy, $V_{\rm circ,sim}=228$~km~s$^{-1}$. Then, we compare the rotation velocity of the maser sources with the gas component of the simulation in Fig.~\ref{snap-fig}.
 
 Interestingly, the rotation velocity of the maser sources (indicated with open stars) show a similar variation in the rotation velocity to our gas components. In the middle panel of $l_{\rm LOS}=120$ deg sample, the maser sources show slower (faster) rotation velocity in the trailing (leading) side, which is consistent with the prediction of the simulation which has a co-rotating spiral arm. Unfortunately, there are insufficient maser sources to test against the simulated co-rotating spiral arm at different radii. However, this shows a possible test of the different explanations for spiral arms.
   
 Fig.~\ref{disvr-fig} shows similar results to Fig.~\ref{disvrot-fig}, but plots the radial (rather than rotational) velocity of stars and gas as a function of the distance from the observer in the simulated galaxy. There are a large variation of Galactocentric radial velocities, $V_{\rm R}$, for star particles observed around the spiral arm, because their epicycle motion is modified owing to the radial migration \citep{gkc14}. Especially in the gas component, the gas particles in the trailing side of the spiral arm are moving outward, and the gas in the leading side of the spiral arm are moving inward. This clear sign of the radial migration is observed at different $l$ and therefore different radii, because of the co-rotating nature of the simulated spiral arm. Again, although the radial velocity of the maser sources show larger scatter in radial velocity, they show a tentative agreement with the simulation results at $l_{\rm LOS}=120$ deg. 

\begin{figure}
\centering
\includegraphics[width=\hsize]{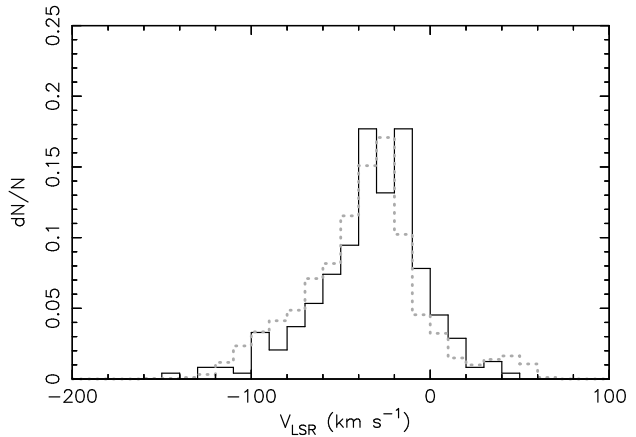}
\caption{
 Distribution of radial velocity from the local standard of the rest for APOGEE DR10 stars (solid histogram) and the star particles (grey dotted histogram) in the simulated galaxy at $(l,b)=(90,0)$ deg. 
}
\label{vlosl90-fig}
\end{figure}

Although currently the the distances for the Milky Way stars are not measured accurately at the distance which we are interested in, $D>$2 kpc, Figs.~\ref{disvrot-fig} and \ref{disvr-fig} indicate that there is a large variation in the velocity for the stars around the spiral arm if the spiral arm is co-rotating. Therefore, it is interesting to compare the Heliocentric radial velocity distribution of observed stars in the Milky Way, which is independent of the errors in the distance, with the simulated star particles. Fig.~\ref{vlosl90-fig} provides a comparison between our simulated data and APOGEE DR10 data \citep{aaaaa14}.

 We have chosen the $(l,b)=(90,0)$ deg data from APOGEE DR10 data, and have applied the same selection as the HQ giant sample of \citet{acsrg14} which is shown in Table 1 of their paper. The solid histogram in Fig.~\ref{vlosl90-fig} shows the distribution of the radial velocity from the local standard of the rest, $V_{\rm LSR}$, for the HQ giant sample at $(l,b)=(90,0)$ deg, where we assume $V_{\odot}=14.6$ km s$^{-1}$ \citep{rmbzd14}. The dotted line is the results for the star particles within $-5<b<5$ deg and $85<l<95$ deg in our simulation. To increase the sample of star particles, we applied a larger selection area for our simulation data than the observed area in APOGEE data. However, for this crude comparison, we do not think that this is a serious issue. Because we do not know the distance of the APOGEE stars, for simplicity we have selected the star particles whose distance from the observer is less than 3.5 kpc, simply assuming that stars further than that are not observed by the APOGEE survey due to the strong extinction in the spiral arm. 

From Fig.~\ref{disvrot-fig}, we expect the variety of $V_{\rm LSR}$ around the co-rotating spiral arm, and it is not surprising to see the large range of $V_{\rm LSR}$ in the simulation data. 
Interestingly, the APOGEE data show a significant number of stars with $V_{\rm LSR}>20$~km~s$^{-1}$, as in the simulations. These high $V_{\rm LSR}$ stars are induced by the co-rotation resonance of the spiral arm in our simulation, and this could be another observational signature of the co-rotating spiral arm. However, \citet{baabbdc12} demonstrated that the Heliocentric radial velocity distribution at $l=90$ deg and the other APOGEE fields can be described well with an axisymmetric model, i.e. the width of the velocity distribution can be explained purely by the velocity dispersion. Therefore, we cannot make any conclusions from this comparison. Still, it is encouraging to see that our simulated galaxy has a velocity distribution consistent with the APOGEE data. 

From Figs.~\ref{disvrot-fig} and \ref{vrothis-fig}, we expect that the velocity distribution would be different between the leading and trailing side of the spiral arm. Therefore, to extract the effect of the spiral arm, the distance information would be crucial. Taking into account stellar populations and the dust extinction may allow meaningful constraints on the distance of the observed stars \citep[e.g.][]{bnrgzc14} and help to identify the location of the resonances by examining the APOGEE data at the different Galactic longitudes. 

\section{Summary}
\label{sum-sec}

We observed our N-body/SPH simulation of a Milky Way-sized disc galaxy in a similar way to how we can observe the Milky Way, with particular interest in the stellar and gas motion around the spiral arm. As our first study, we have focused on the three lines of sight on the disc plane, $(l,b)=(90,0), (120,0)$ and (150,0) deg, and analysed the rotation and radial velocity of stars and gas as a function of the distance from our assumed location of the observer.  Similarly to the recent literature based on N-body simulations, our simulated galaxy shows a co-rotating spiral arm, i.e. the spiral arm is rotating with the same speed as the circular velocity, at the lines of sight selected. We show that the stars around the spiral arm show a large variation in both radial and rotational velocities owing to the co-rotating spiral arm. If the spiral arm is indeed a co-rotating spiral arm, we should observe a similar variation of the rotation and radial velocities around the spiral arm at every Galactocentric radius. An accurate measurement of the distance, proper motion and radial velocity of the stars is required, and the {\it Gaia} data will be a critical test for the co-rotating spiral arm. 

 We show that the stars behind the spiral arm always gain angular momentum, while the stars at the front of the spiral arm lose angular momentum. The stars tend to rotate slower (faster) behind (at the front of) the spiral arm and move outward (inward). Because of the epicycle motion of the stars, we also see the stars with high (low) rotation velocity behind (at the front of) the spiral arms. We find that these stars came from the outer (inner) region and decelerated (accelerated) at the front of (behind) the arm. Then, they are passed by (passed) the spiral arm and are observed at their peri-centre (apo-centre) phase. These are consistent with \citet{gkc14}, and indicate a variety of orbits owing to the co-rotating spiral arm. These variety of orbits are likely to be closely related to the formation and disruption of the spiral arm. However, we need further investigation to reach a firm conclusion. Still, this study indicates that numerical simulations provides useful tests in the study of the nature of spiral arms.
 
 We have also analysed the rotation and radial velocity of the cold gas component. This is much simpler than the stars. We found a clear trend in the gas component which rotates slower (faster) and moves outward (inward) behind (at the front of) the spiral arm. We have compared the results with the observed data of the maser sources from \citet{rmbzd14}. Interestingly, the data show similar trend in $V_{\rm rot}$ around the expected position of the spiral arm around $l=120$ deg. More data from the accurate astrometric measurement of the maser sources will provide additional constraints on the nature of the spiral arm. 
 
 Although we have observed three lines of sight for one snapshot of the numerical simulation, the observed gas and stellar motions are naturally expected features from our previous studies \citep{gkc12a,gkc12b,gkc14}, and therefore the observed trend should be common features in the co-rotating spiral arms in N-body/SPH simulations. We have analyzed several snapshots at different timesteps of the simulation, and confirmed that similar kinematic trends are always observed.
 
 Encouraged by the success of this study, we are currently working to make a quantitative prediction for the upcoming {\it Gaia} data for the co-rotating spiral arm, by taking into account the stellar population, dust extinction and expected {\it Gaia} errors by improving the methods in \citet{pck12} and \citet{hk14}. The entire topic will be further illuminated by another set of theoretical models, e.g. the analytical model of the stellar kinematics from the spiral arm, and also the numerical simulations with a fixed spiral arm potential and a constant pattern speed including both gas and stars \citep[e.g.][]{wbs11,dpn14}.
      
 \section*{Acknowledgments}
We thank an anonymous referee for their constructive comments and helpful suggestions which have improved the manuscript. We also thank Jo Bovy and the anonymous referee for their critical comments on our over-interpretation of Fig.~\ref{vlosl90-fig} in the first version of our draft.
The calculations for this paper were performed on the UCL facility Legion, the IRIDIS HPC facility provided by the Centre for Innovation and the DiRAC Facilities (through the COSMOS consortium) jointly funded by STFC and the Large Facilities Capital Fund of BIS.   
We also acknowledge PRACE for awarding us access to resource Cartesius based in Netherlands at SURFsara.

Funding for SDSS-III has been provided by the Alfred P. Sloan Foundation, the Participating Institutions, the National Science Foundation, and the U.S. Department of Energy Office of Science. The SDSS-III web site is http://www.sdss3.org/.

SDSS-III is managed by the Astrophysical Research Consortium for the Participating Institutions of the SDSS-III Collaboration including the University of Arizona, the Brazilian Participation Group, Brookhaven National Laboratory, Carnegie Mellon University, University of Florida, the French Participation Group, the German Participation Group, Harvard University, the Instituto de Astrofisica de Canarias, the Michigan State/Notre Dame/JINA Participation Group, Johns Hopkins University, Lawrence Berkeley National Laboratory, Max Planck Institute for Astrophysics, Max Planck Institute for Extraterrestrial Physics, New Mexico State University, New York University, Ohio State University, Pennsylvania State University, University of Portsmouth, Princeton University, the Spanish Participation Group, University of Tokyo, University of Utah, Vanderbilt University, University of Virginia, University of Washington, and Yale University.
 
\bibliographystyle{mn}
%\bibliography{../dkref}

\label{lastpage}

\end{document}